\pdfoutput=1
\documentclass{article}
\usepackage[utf8]{inputenc}
\usepackage[english]{babel}
\usepackage{hyperref}
\usepackage{amsmath}
\usepackage{amssymb}
\usepackage{booktabs}
\usepackage{rotating}
\usepackage{titling}
\usepackage{titlesec}
\usepackage{graphicx}
\usepackage{float}
\usepackage{tablefootnote}
\usepackage{textcomp}
\usepackage{array}
\usepackage{hyperref}
\usepackage{setspace}
\usepackage{pdfpages}
\usepackage{import}
\usepackage{xr}
\newcommand{\ra}[1]{\renewcommand{\arraystretch}{#1}}

\DeclareMathOperator{\relu}{ReLU}
\DeclareMathOperator{\softplus}{Softplus}
\DeclareMathOperator{\softmax}{Softmax}
\DeclareMathOperator{\crossentropy}{CE}
\usepackage{xcolor}

\usepackage[style=nature, citestyle=numeric, sorting=none, uniquename=false]{biblatex}
\DeclareLanguageMapping{english}{english-apa}
\DeclareFieldFormat{labelnumberwidth}{\mkbibbrackets{#1}}

\usepackage{geometry}
\defbibenvironment{bibliography}
  {\list
     {\printtext[labelnumberwidth]{%
      \printfield{labelprefix}%
      \printfield{labelnumber}}}
     {\setlength{\labelwidth}{\labelnumberwidth}%
      \setlength{\leftmargin}{\labelwidth}%
      \setlength{\labelsep}{\biblabelsep}%
      \addtolength{\leftmargin}{\labelsep}%
      \setlength{\itemsep}{\bibitemsep}%
      \setlength{\parsep}{\bibparsep}}%
      }
  {\endlist}
  {\item}

\begin{document}
\begin{center}
{\Large \textbf{Deep Representation Learning of Electronic Health Records \\ to Unlock Patient Stratification at Scale}}
\end{center}
\bigskip

\begin{center}
Isotta Landi$^{1,2}$, Benjamin S. Glicksberg$^{3,4,5}$, Hao-Chih Lee$^{4,5}$, Sarah Cherng$^{4,5}$, Giulia Landi$^6$, Matteo Danieletto$^{3,4,5}$, Joel T. Dudley$^{4,5}$, Cesare Furlanello$^{\S,1, 7}$, and Riccardo Miotto$^{*,\S,3,4,5}$
\end{center}
\begin{flushleft}
\small{\S\ These authors share senior authorship.}
\end{flushleft}
\bigskip

\begin{center}
(1) Bruno Kessler Institute\\
Via Sommarive 18, 38123 Povo (TN), Italy\\
(2) Department of Psychology and Cognitive Science\\
University of Trento\\
Corso Bettini 84, 38068 Rovereto (TN), Italy\\
(3) Hasso Plattner Institute for Digital Health at Mount Sinai\\
(4) Institute for Next Generation Healthcare\\
(5) Department of Genetics and Genomic Sciences\\
Icahn School of Medicine at Mount Sinai\\
1 Gustave L. Levy Place, New York, NY 10029, USA\\
(6) Department of Mental Health and Pathological Addiction\\
Azienda USL Centro ``Santi''\\
Via Vasari 13, 43100 Parma, Italy\\
(7) HK3 Lab\\
Via Castel Morrone 14, 20129 Milan, Italy
\end{center}
\vspace*{\fill}
\begin{flushleft}
\textbf{Corresponding author:}\\
Riccardo Miotto, PhD\\
Hasso Plattner Institute for Digital Health at Mount Sinai\\
Department of Genetics and Genomic Sciences\\
Icahn School of Medicine at Mount Sinai\\
1 Gustave L. Levy Place\\
New York, NY 10029 \\
USA\\
email: riccardo.miotto@mssm.edu
\end{flushleft}
\newpage
 
\begin{abstract}
Deriving disease subtypes from electronic health records (EHRs) can guide next-generation personalized medicine. However, challenges in summarizing and representing patient data prevent widespread practice of scalable EHR-based stratification analysis. Here we present an unsupervised framework based on deep learning to process heterogeneous EHRs and derive patient representations that can efficiently and effectively enable patient stratification at scale.
We considered EHRs of $1,608,741$ patients from a diverse hospital cohort comprising of a total of $57,464$ clinical concepts. We introduce a representation learning model based on word embeddings, convolutional neural networks, and autoencoders (i.e., ConvAE) to transform patient trajectories into low-dimensional latent vectors. We evaluated these representations as broadly enabling patient stratification by applying hierarchical clustering to different multi-disease and disease-specific patient cohorts.
ConvAE significantly outperformed several baselines in a clustering task to identify patients with different complex conditions, with $2.61$ entropy and $0.31$ purity average scores. When applied to stratify patients within a certain condition, ConvAE led to various clinically relevant subtypes for different disorders, including type 2 diabetes, Parkinson's disease and Alzheimer's disease, largely related to comorbidities, disease progression, and symptom severity.
With these results, we demonstrate that ConvAE can generate patient representations that lead to clinically meaningful insights. This scalable framework can help better understand varying etiologies in heterogeneous sub-populations and unlock patterns for EHR-based research in the realm of personalized medicine.

\end{abstract}
\newpage

\doublespacing

\section*{Introduction}

Electronic health records (EHRs) are collected as part of routine care across the vast majority of healthcare institutions. They consist of heterogeneous structured and unstructured data elements, including demographic information, diagnoses, laboratory results, medication prescriptions, free text clinical notes, and images. EHRs provide snapshots of a patient's state of health and have created unprecedented opportunities to investigate the properties of clinical events across large populations using data-driven approaches and machine learning. At the individual level, patient trajectories can foster personalized medicine; across a population, EHRs can provide a vital resource to understand population health management and help make better decisions for healthcare operation policies \cite{jensen2012}.

Personalized medicine focuses on the use of patient-specific data to tailor treatment to an individual's unique health characteristics. However, even seemingly simple diseases can show different degrees of complexity that can create challenges for identification, treatment, and prognosis, despite equivalence at the diagnostic level \cite{cutting2015, alexandrov2016}. Heterogeneity among patients is particularly evident for \emph{complex disorders}, where the etiology is due to an amalgamation of multiple genetic, environmental, and lifestyle factors. Several different conditions have been referred to as \emph{complex}, such as Parkinson's disease (PD) \cite{langston2006}, multiple myeloma (MM) \cite{mel2014}, and type 2 diabetes (T2D) \cite{pearson2019}. Patients with complex disorders may differ on multiple systemic layers (e.g., different clinical measurements or comorbidity landscape) and in response to treatments, making these conditions difficult to model. Multiple data types in patient longitudinal EHR histories offer a way to examine disease complexity and present an opportunity to refine diseases into subtypes and tailor personalized treatments. This task is usually referred to as ``EHR-based patient stratification''. This follows a common approach in clinical research, where attempts to identify latent patterns within a cohort of patients can contribute to the development of improved personalized therapies \cite{dugger2017}. 

From a computational perspective, patient stratification is a data-driven, unsupervised learning task that groups patients according to their clinical characteristics \cite{baytas2017patient}. Previous work in this domain aggregates clinical data at a patient level, representing each patient as multi-dimensional vectors, and derives subtypes within a disease-specific population via clustering (e.g., in autism \cite{doshivelez2013}) or topological analysis (e.g., for T2D \cite{li2015identification}). Deep learning has been applied to derive more robust patient representations to improve disease subtyping \cite{baytas2017patient, zhang2019data}. Baytas et al. used time-aware long short-term memory (LSTM) networks to leverage stratification of longitudinal data of PD patients \cite{baytas2017patient}. Similarly, Zhang et al. used LSTM to identify three subgroups of patients with idiopathic PD that differ in disease progression patterns and symptom severity \cite{zhang2019data}. These studies, however, only focused on curated and small disease-specific cohorts, with ad hoc manually selected features. This approach not only limits scalability and generalizability, but also hinders the possibility to discover unknown patterns that might characterize a condition. Because EHRs tend to be incomplete, using a diverse cohort of patients to derive disease-specific subgroups can adequately capture the features of heterogeneity within the disease of interest \cite{chen2019deep}. However, it is challenging to create large-scale computational models from EHRs because of data quality issues, such as high dimensionality, heterogeneity, sparseness, random errors, and systematic biases. Advances in machine learning, specifically in representation learning \cite{bengio2013representation} and deep learning \cite{lecun2015deep}, are introducing different computational models to leverage EHRs for personalized healthcare \cite{miotto17review, xiao2018review}. This work fits into this landscape by presenting an unsupervised patient stratification pipeline that aims to automatically detect clinically meaningful subtypes within any condition by using patient representations learned from a heterogeneous and large cohort of EHRs.

\begin{figure}
 \centering
 \begin{tabular}{c}
 \multicolumn{1}{l}{
 \textbf{a}}\\
   \includegraphics[width=\textwidth]{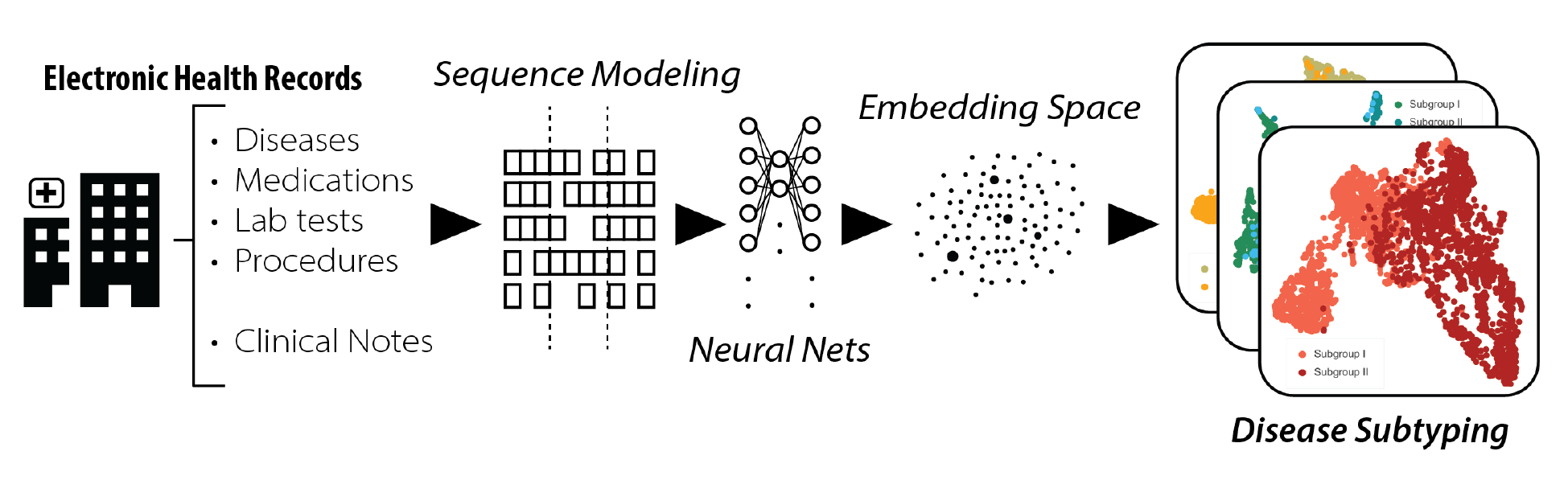}\\
 [3pt]
 \multicolumn{1}{l}{
 \textbf{b}}\\
  \includegraphics[width=\textwidth]{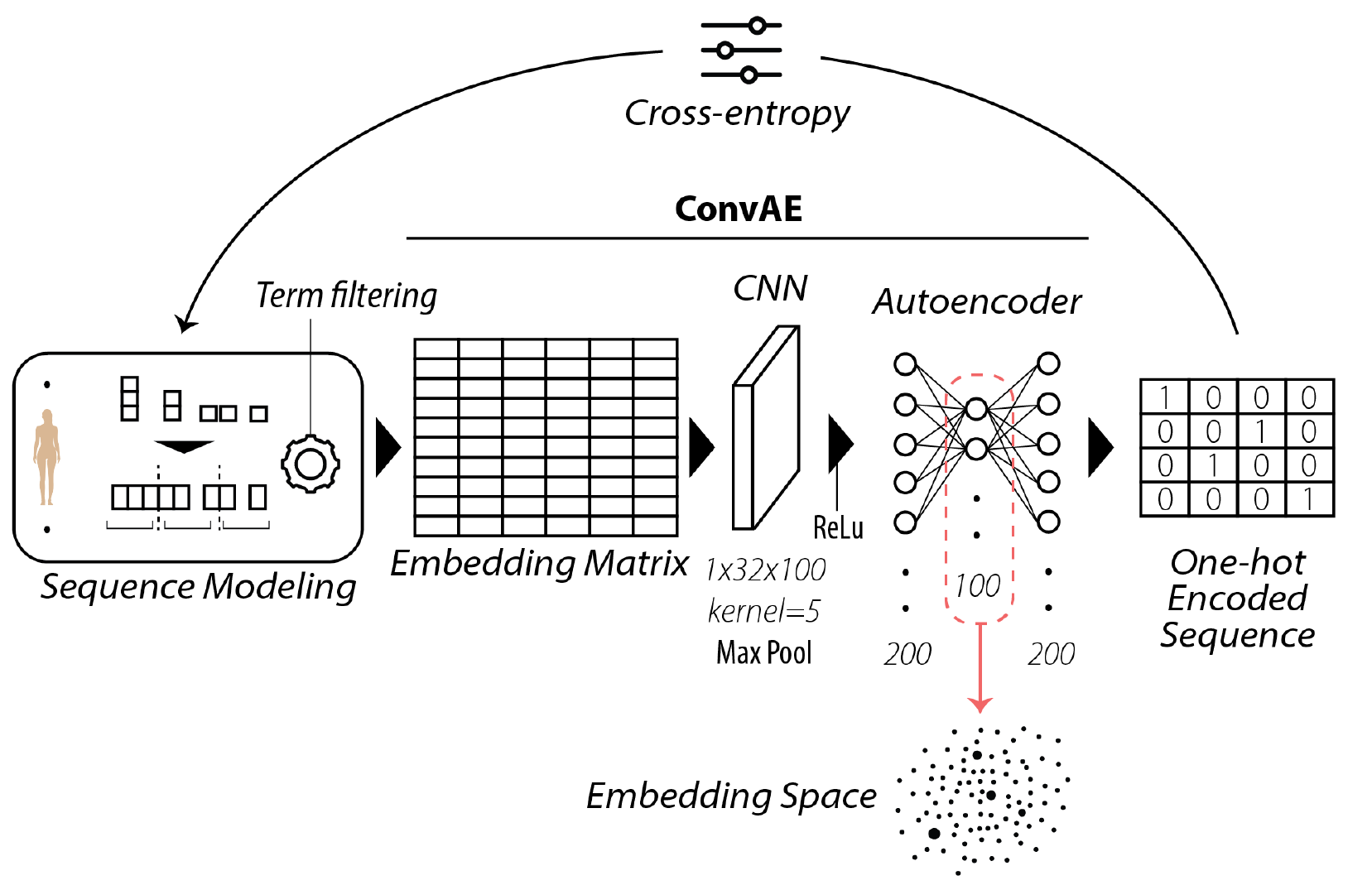}\\
 [3pt]
  \end{tabular}
\caption{Patient stratification framework and ConvAE architecture. (\textbf{a}) Framework enabling patient stratification analysis from deep unsupervised EHR representations; (\textbf{b}) Details of the ConvAE representation learning architecture.}
 \label{fig:arch}
 \end{figure}

In particular, this paper proposes a general framework for identifying disease subtypes at scale (see Figure~\ref*{fig:arch}a). We first propose an unsupervised deep learning architecture to derive vector-based patient representations from a large and domain-free collection of EHRs. This model (i.e., ConvAE) combines 1) embeddings to contextualize medical concepts, 2) convolutional neural networks (CNNs) to loosely model the temporal aspects of patient data, and 3) autoencoders (AEs) to enable the application of an unsupervised architecture. Second, we show that ConvAE-based representations learned from real-world EHRs of about $1.6$M patients from the Mount Sinai Health System in New York improve clustering of patients with different disorders compared to several commonly used baselines. Last, we demonstrate that ConvAE leads to effective patient stratification with minimal effort. To this end, we used the encodings learned from domain-free and heterogeneous EHRs to derive subtypes for different complex disorders and provide a qualitative analysis to determine their clinical relevance.

This architecture enables patient stratification at scale by eliminating the need for manual feature engineering and explicit labeling of events within patient care timelines, and processes the whole EHR sequence regardless of the length of patient history. By generating disease subgroups from large-scale EHR data, this architecture can help disentangle clinical heterogeneity and identify high-impact patterns within complex disorders, whose effect may be masked in case-control studies \cite{manchia2013}. The specific properties of the different subgroups can then potentially inform personalized treatments and improve patient care.

\section*{Results}

We first evaluated the extent to which ConvAE-based patient representations can be used to identify different clinical diagnoses in the EHRs (i.e., disease phenotyping \cite{shah18}). To this end, we performed clustering analysis using patients with the following eight complex disorders: T2D, MM, PD, Alzheimer's disease (AD), Crohn's disease (CD), breast cancer (BC), prostate cancer (PC), and attention deficit hyperactivity disorder (ADHD). We used SNOMED--CT (Systematized nomenclature of medicine -- clinical terms) \cite{cote1980} to find all patients in the data warehouse diagnosed with these conditions; see Supplementary Table \ref*{tab:disclass} and the ``Multi-disease clustering analysis'' subsection in ``Methods'' for more details.

Evaluation was organized as a 2-fold cross-validation experiment to show model generalizability and to assess replication of the stratification results. To this aim, we randomly split the dataset in half, obtaining two independent cohorts of about $800,000$ patients that we used to train and test the models (and vice versa). While we used all patients in each cohort for training, in the test sets we retained only the patients diagnosed with one of the eight disorders under consideration, obtaining about $94,000$ test patients per fold (see the ``Dataset'' subsection in ``Methods'' for more details). 

Table \ref*{tab:result} shows the results using hierarchical clustering for different ConvAE architectures (one, two, and multikernel CNN layers) and baselines in terms of entropy and purity scores averaged over the 2-fold cross-validation experiment. ConvAE performed significantly better than other models largely used in healthcare for representation learning, including Deep Patient \cite{miotto2016deep}, for both entropy and purity scores ($p_s<0.001$, t-tests comparisons with Bonferroni correction). The configuration with one CNN layer yielded the best overall performance and the learned encodings produced clusters associated with the largest number of distinct diseases (i.e, $6.50$, based on purity score analysis). It is worth saying that, without a predictive theory of clustering \cite{dougherty2004probabilistic, dalton2018optimal}, validation metrics frequently fail to correlate with clustering errors \cite{brun2007model}. However, such theoretic structure is not applicable in this context because the heterogeneity of the external complex disorder classes do not provide a reliable probabilistic framework. For this reason, we used, rather than estimation error analysis, transparent external metrics, such as entropy and purity scores, which evaluate cluster composition and also account for possible subgroups of complex diseases \cite{amigo2009comparison}.

\begin{figure}
 {\centering
\includegraphics[width=\textwidth]{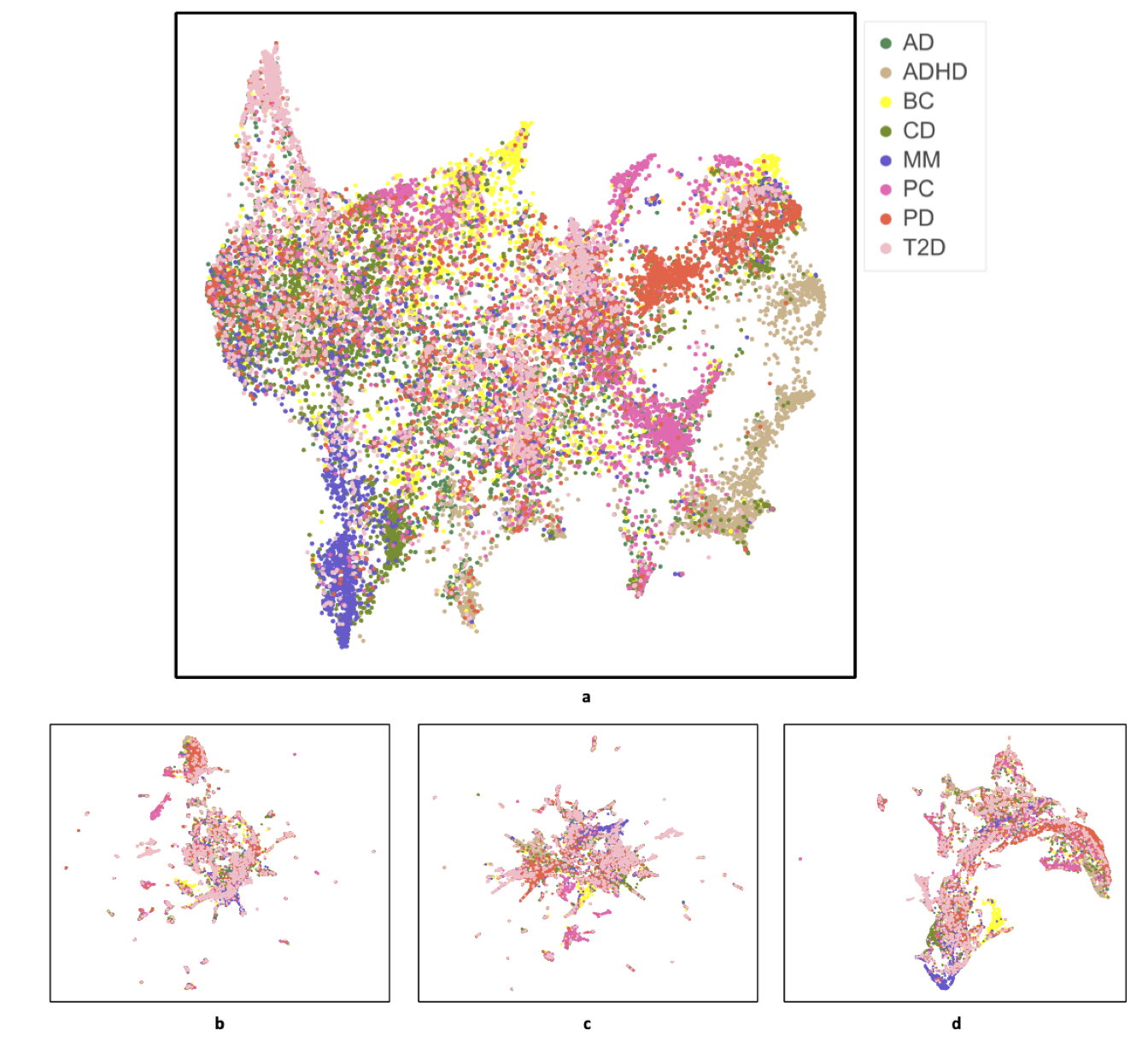}}
{\footnotesize AD = Alzheimer's disease; ADHD = Attention deficit hyperactivity disorder; BC = Breast cancer; CD = Crohn's disease;\\ MM = Multiple myeloma; PC = Prostate cancer; PD = Parkinson's disease; T2D = Type 2 diabetes}
 \caption{Uniform manifold approximation and projection (UMAP) encoding visualization. (\textbf{a}) ConvAE 1-layer CNN; (\textbf{b}) SVD-RawCount; (\textbf{c}) SVD-TFIDF; (\textbf{d}) Deep Patient. AD = Alzheimer's disease; ADHD = Attention deficit hyperactivity disorder; BC = Breast cancer; CD = Crohn's disease; MM = Multiple myeloma; PC = Prostate cancer; PD = Parkinson's disease; T2D = Type 2 diabetes.}
 \label{fig:encodings}
 \end{figure}

\begin{figure}
 {\centering
 \includegraphics[width=\textwidth]{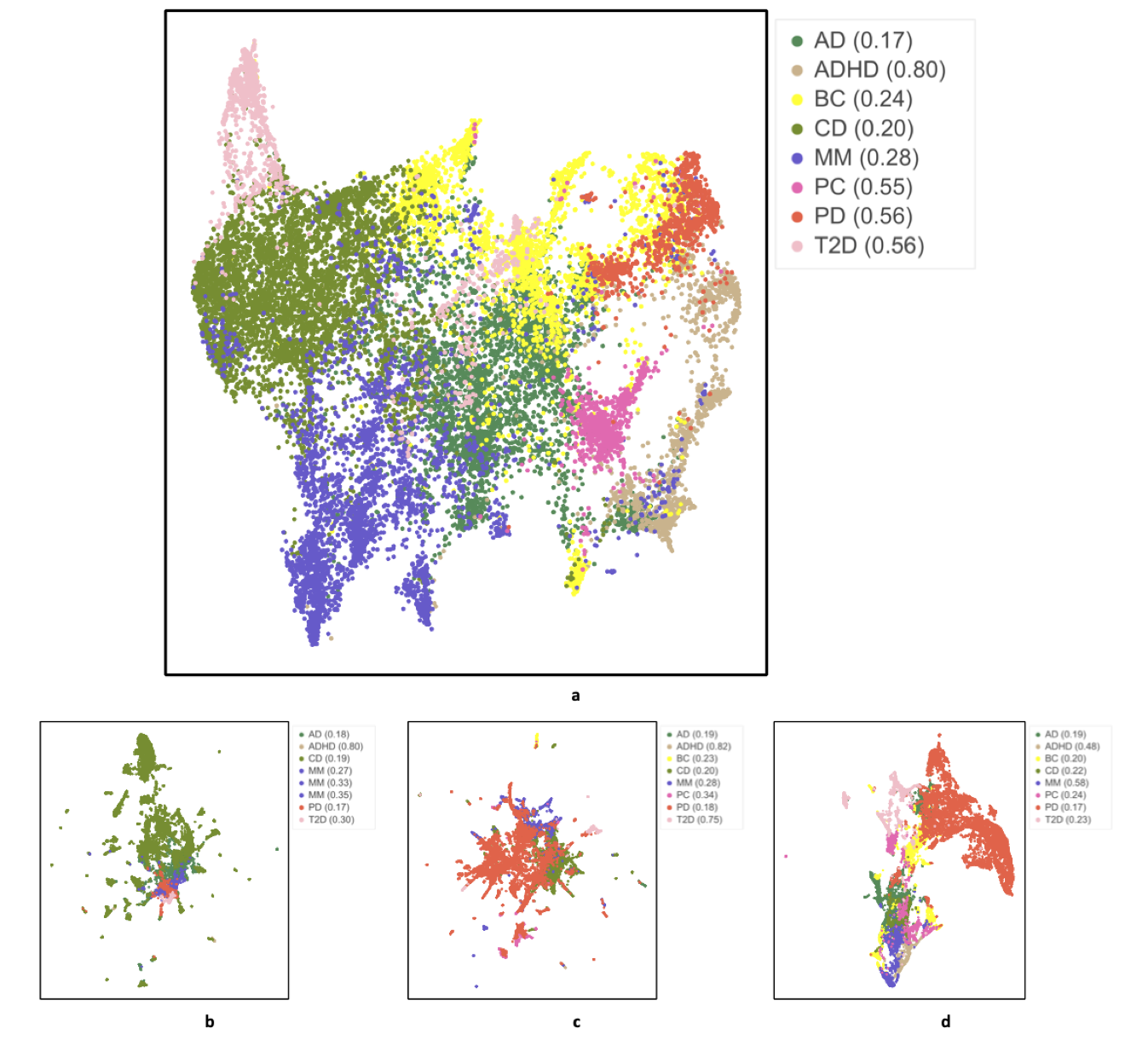}}
{\footnotesize AD = Alzheimer's disease; ADHD = Attention deficit hyperactivity disorder; BC = Breast cancer; CD = Crohn's disease;\\ MM = Multiple myeloma; PC = Prostate cancer; PD = Parkinson's disease; T2D = Type 2 diabetes}
\caption{Uniform manifold approximation and projection (UMAP) clustering visualization. (\textbf{a}) ConvAE 1-layer CNN; (\textbf{b}) SVD-RawCount; (\textbf{c}) SVD-TFIDF; (\textbf{d}) Deep Patient. AD = Alzheimer's disease; ADHD = Attention deficit hyperactivity disorder; BC = Breast cancer; CD = Crohn's disease; MM = Multiple myeloma; PC = Prostate cancer; PD = Parkinson's disease; T2D = Type 2 diabetes.}
 \label{fig:clustering}
 \end{figure}

Figure \ref*{fig:encodings} visualizes the distribution of the different patient representations along with their disease cohort labels obtained using UMAP (Uniform manifold approximation and projection for dimension reduction \cite{mcinnes2018umap}). ConvAE captures hidden patterns of overlapping phenotypes while still displaying identifiable groups of patients with distinct disorders. Figure~\ref*{fig:clustering} shows the same patient distribution highlighting clustering labels and purity percentage scores of each cluster dominating disease. These figures refer to only one of the cross-validation splits; results for the second split are similar and are available in Supplementary Figures \ref*{fig:encodings2} and \ref*{fig:clustering2}). ConvAE (with one CNN layer) also led to better clustering, visually, than all baselines. Patients with ADHD were the most separated and detected with $80\%$ purity by hierarchical clustering. Visible clusters with $> 50\%$ purity were also identified for T2D, PC and PD. Comparing the encoding projections (Figure~\ref*{fig:encodings}) to the clustering visualization (Figure~\ref*{fig:clustering}), we observe that patients whose disease is not correctly identified by clusters tend to not clearly separate in this low-dimensional space. As an example, AD patients were randomly scattered in the plot and did not lead to distinguishable clusters. This might be due to factors such as sex and age, intrinsic biases or noise, but it might also reflect a shared phenotypic characterization that drives the learning process into displaying these patient EHR progressions closely together irrespective of disease labels. 

\begin{figure}
 \centering
  \includegraphics[width=.9\textwidth, height=.9\textheight, keepaspectratio]{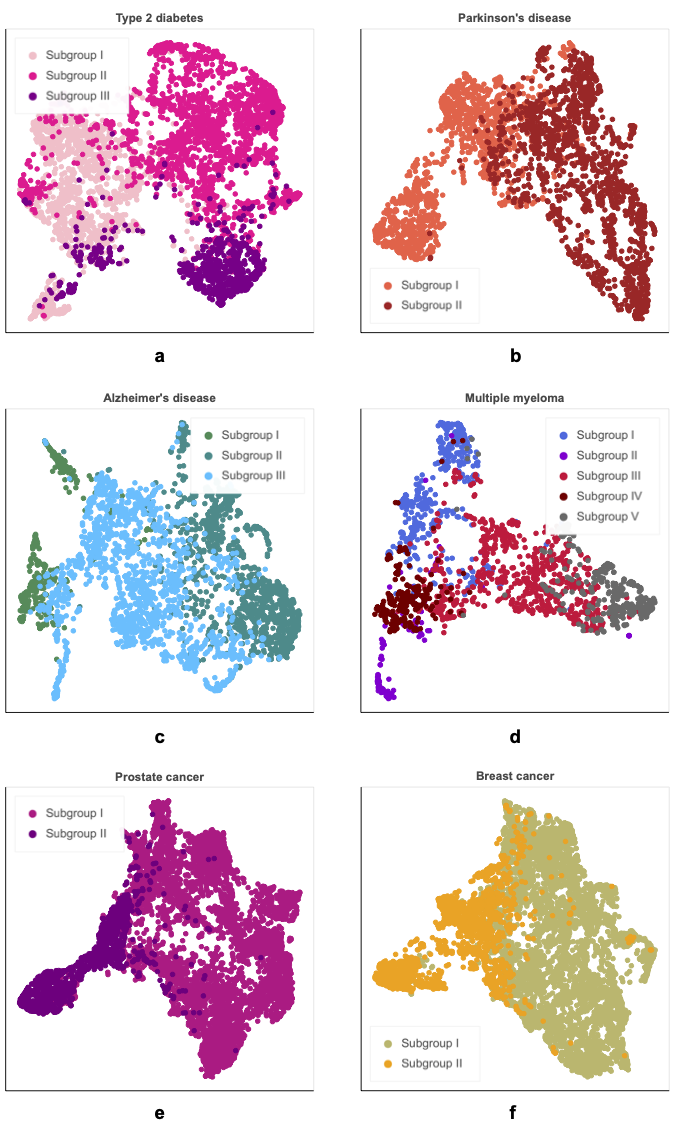}
  \caption{Complex disorder subgroups. A subsample of $5,000$ patients with T2D is displayed in Figure (\textbf{a}). Figures 
 (\textbf{b}), (\textbf{c}), (\textbf{d}), (\textbf{e}), (\textbf{f}) display patient subtypes for Parkinson's and Alzheimer's disease, multiple myeloma, prostate and breast cancer cohorts, respectively.}
 \label{fig:innerval}
 \end{figure}

We then evaluated the use of ConvAE representations for patient stratification at scale and the identification of clinically relevant disease subtypes. We considered six diseases: T2D, PD, AD, MM, PC, and BC. These are all age-related complex disorders with late onset (i.e., averaged increased prevalence after $60$ years of age) \cite{cowie2018, lau2006, qiu2009, kazandjian2016, pc1, bc2}. We decided to focus on these conditions to avoid, to some extent, the confounding effect of age that could affect learning and the evaluation of different subtypes. Figure \ref*{fig:innerval} shows results running hierarchical clustering on the ConvAE-based patient representations of each different disease cohort. To determine the optimal number of clusters, we empirically selected the smallest number of clusters that minimize the increase in explained variance (i.e., Elbow method). We were able to identify different subtypes for each disease with no additional feature selection and using representations derived from a domain-free cohort of patients. Supplementary Table~\ref*{tab:inner} reports the number of patients in each cohort and the number of subgroups identified. Similar results were obtained for the second split and are reported in Supplementary Figure~\ref*{fig:innerval2}. 

In the following sections, we present the clinical characterization of T2D, PD, and AD subgroups via enrichment analysis of medical concept occurrences (see Supplementary Material for the characterization of the other conditions). We compare T2D and PD results to related studies based on ad hoc cohorts \cite{li2015identification, zhang2019data}. Conversely, there are no published EHR-based stratification studies for AD, MM, PC, and BC to use for comparison. All subtypes were reviewed by a clinical expert to highlight meaningful descriptors and we used multiple pairwise chi-squared tests to assess group differences. For each disease, we list sex and age statistics of the cohort (between group comparisons are performed via multiple pairwise chi-squared tests and t-tests), as well as the five most frequent diagnosis, medications, laboratory tests, and procedures, ordered according to in-group and total frequencies, in Supplementary Tables~\ref*{tab:t2d}-\ref*{tab:bc}. The results for the second split are reported in Supplementary Tables~\ref*{tab:t2d2}-\ref*{tab:bc2}.
 
\subsection*{Type 2 diabetes}

Patients with T2D clustered into three different subgroups that relate to different stages of progression for the disease (see Figure~\ref*{fig:innerval}a and Supplementary Table \ref*{tab:t2d} for details).

Subgroup I included $18,325$ patients and represents the mild symptom severity cohort, characterized by common T2D symptoms (e.g., metabolic syndrome), which were treated with \emph{Metformin}, an oral hypoglycemic medication. Moreover, it also included patients exposed to lifestyle risk factors, such as \emph{Obesity} \cite{pearson2019}.

Subgroups II/III, which were composed by $22,659$ and $7,704$ patients, respectively, showed concomitant conditions associated to T2D progression and worsening symptoms. Specifically, subgroup II clustered patients characterized by microvascular problems, such as diabetic nephropathy, neuropathy, and/or peripheral artery disease. The significant presence of \emph{Creatinine} and \emph{Urea nitrogen} laboratory tests, which estimate renal function, suggests monitoring of kidney diseases, which are often related to T2D \cite{vallon2011}. The presence of \emph{Pain in limb}, combined with analgesic drugs (i.e., \emph{Paracetamol}, \emph{Oxycodone}), indicates the presence of vascular lesions at the peripheral level, manifested as ischemic rest pain or ulceration. This was confirmed by \emph{Peripheral vascular disease} diagnoses which accounts for $50\%$ of terms in the T2D cohort.

Subgroup III showed severe cardiovascular problems, identified by a significant presence of medical concepts related to coronary artery diseases, e.g., \emph{Coronary atherosclerosis}, \emph{Angina pectoris}, which are serious risk factors for heart failure. These subjects were often treated with antiplatelet therapy (i.e., \emph{Acetylsalicylic acid, Clopidrogel}) to prevent cardiovascular events (e.g., stroke) and were likely to receive invasive procedures to treat severe arteriopathy. For instance, $30\%$ of patients in subgroup III underwent \emph{Percutaneous Transluminal Coronary Angioplasty}, a procedure to open up blocked coronary arteries. 

Our results confirm, in part, what was observed by Li et al. \cite{li2015identification}, which used topology analysis on an ad hoc cohort of T2D patients and identified three distinct subgroups characterized by 1) microvascular diabetic complications (i.e., diabetic nephropathy, diabetic retinopathy); 2) cancer of bronchus and lungs; and 3) cardiovascular diseases and psychiatric disorders. In particular, we detected the same microvascular and cardiovascular disease groups, which are consequences of T2D. In contrast, we were unable to detect a subgroup significantly characterized by cancer, an epiphenomenon that can be caused by secondary immunodeficiency in patients with T2D \cite{malaguarnera2010, delamaire1997}. See Supplementary Material for further description and a clustering comparison via Fowlkes-Mallows index.

\subsection*{Parkinson's disease}

Individuals diagnosed with PD divided into two groups (Figure~\ref*{fig:innerval}b and Supplementary Table \ref*{tab:pd}): one dominated by motor symptoms ($1,368$ patients) and another ($1,684$ patients) characterized by non-motor/independent features and longer course of disease. 

Subgroup I is characterized as a tremor-dominant cohort (i.e., manifested by motor symptoms) because of the significant presence of diagnosis such as \emph{Essential tremor}, \emph{Anxiety state}, and \emph{Dystonia}. It is interesting to note that motor clinical features likely led to a common misdiagnosis of essential tremor, which is an action tremor that typically involves the hands. Parkinsonian tremor, on the contrary, although can be present during postural maneuvers and action, is much more severe at rest and decreases with purposeful activities. However, when the tremor is severe, it is difficult to distinguish action tremor from resting tremor, leading to the aforementioned misdiagnosis \cite{jain2006}. Moreover, anxiety states, emotional excitement, and stressful situations can exacerbate the tremor, and lead to a delayed PD diagnosis. \emph{Brain MRI}, usually non-diagnostic in PD, was ordered for several patients in this subgroup ($13\%$) suggesting its use for differential diagnosis, e.g., to investigate the presence of chronic/vascular encephalopathy. 

Subgroup II included non-motor and independent symptoms, such as \emph{Constipation} and \emph{Fatigue}. Patients in subgroup II were significantly diagnosed with \emph{Coronary artery disease} that is prevalent in older patients ($> 50$ years old). Constipation and fatigue are among the most common non-motor problems related to autonomic dysfunction, diminished activity level, and slowed intestinal transit time in PD \cite{alves2004, siciliano2018}.

In their study about PD stratification with PPMI (Parkinson's progression markers initiative) data, Zhang et al. \cite{zhang2019data} identified three distinct subgroups of patients based on severity of both motor and non-motor symptoms. In particular, one subgroup included patients with moderate functional decay in motor ability and stable cognitive ability; a second subgroup presented with mild functional decay in both motor and non-motor symptoms; and the third subgroup was characterized by rapid progression of both motor and non-motor symptoms. EHRs do not quantitatively capture PD symptom severity, therefore our analyses cannot replicate these findings. However, unlike Zhang et al., we can discriminate between specific motor and non-motor symptoms and also suggest a longer, but not necessarily more severe, disease course for the non-motor symptom subgroup.

\subsection*{Alzheimer's disease}

Patients with AD separated into three subgroups marked by AD onset, disease progression, and severity of cognitive impairment (see Figure~\ref*{fig:innerval}c and Supplementary Table \ref*{tab:ad}).

Subgroup I is characterized by $399$ patients with early-onset AD, i.e., patients whose dementia symptoms have typically developed between the age of $30$ and $60$ years, and initial neurocognitive disorder. Early-onset AD affects $5\%$ of the individuals with AD in the US \cite{ad} and, because clinicians do not usually look for AD in younger patients, the diagnostic process includes extensive evaluations of patient symptoms. In particular, given that a certain AD diagnosis can only be provided post-mortem through brain examination, clinicians first rule out other causes that can lead to early-onset dementia (i.e., differential diagnosis). We find evidence of this practice in this subgroup, which includes postmenopausal women, identifiable by mean age greater than $50$, \emph{Osteoporosis} diagnosis with calcium supplement therapy, and menopausal hormone treatment (i.e., \emph{Estradiol}). Patients in this group are also tested for infectious diseases (e.g., HIV, Syphilis, Hepatitis C, Chlamydia/Gonorrhoea) that are possible causes of early-onset dementia \cite{manji2013}, and screened via structural neuroimaging, e.g., \emph{MRI/PET brain}. As cognitive dysfunctions that may be mistaken for dementia can also be caused by depression and other psychiatric conditions, the presence of \emph{Psychiatric service/procedure} suggests psychiatric evaluations to exclude depressive pseudodementia. After the differential diagnosis process and the exclusion of other possible causes, eventually these patients received a diagnosis of AD. 

Subgroup II includes $1,170$ patients with late-onset AD, mild neuropsychiatric symptoms and cerebrovascular disease. Here, the absence of behavioral disturbances in $39\%$ of patients, and their high average age ($M=84.96,\ sd=9.61$) suggest a late AD onset, with a progression characterized by a slower rate of cognitive ability decline \cite{lyketsos2002}. Moreover, the presence of \emph{Acetylsalicylic acid}, an antiplatelet medication, and \emph{Intracranial hemorrage} diagnosis indicates the co-occurrence of cerebrovascular disease, which affects blood vessels and blood supply to the brain. Cerebrovascular diseases are common in aging, and can often be associated with AD \cite{snyder2015}. In this regard, \emph{Head CT} may have been performed to prevent or identify structural abnormalities related to cerebrovascular disease. 

Subgroup III is characterized by $1,632$ individuals with typical onset and mild-to-moderate dementia symptoms. A cohort of $409$ patients was treated with \emph{Donepezil}, a cholinesterase inhibitor, that is a primary treatment for cognitive symptoms and it is usually administered to patients with mild-to-moderate AD, producing small improvement in cognition, neuropsychiatric symptoms, and activities of daily living \cite{birks2018}. Patients in this subgroup also showed both dementia with and without behavioral disturbances.

\section*{Discussion}

This study proposes a computational framework to disentangle the heterogeneity of complex disorders in large-scale EHRs through the identification of data-driven clinical patterns with machine learning. Specifically, we developed and validated an unsupervised architecture based on deep learning (i.e., ConvAE) to infer informative vector-based representations of millions of patients from a large and diverse hospital setting, which facilitates the identification of disease subgroups that can be leveraged to personalize medicine. These representations aim to be domain-free (i.e., not related to any specific task since learned over a large multi-domain dataset) and enable patient stratification at scale. Results from our experiments show that ConvAE significantly outperformed several baselines on clustering patients with different complex conditions and led to the identification of different clinically meaningfully disease subtypes.

Results identified disease progression, symptom severity, and comorbidities as contributing the most to the EHR-based clinical phenotypic variability of complex disorders. In particular, T2D patients divided into three subgroups according to comorbidities (i.e., cardiovascular and microvascular problems) and symptom severity (i.e., newly diagnosed with milder symptoms). Individuals with PD showed different disease duration and symptoms (i.e., motor, non-motor). AD profiles distinguished early- and late-onset groups and separate patients with mild neuropsychiatric symptoms and cerebrovascular disease from patients with mild-to-moderate dementia. Patients with MM were characterized by different comorbidities (e.g., amyloidosis, pulmonary diseases) that manifest alongside precise typical signs of MM. Patients with PC and BC separated according to disease progression. These findings showed that the features learned by ConvAE describe patients in a way that is general and conducive to identifying meaningful insights into different clinical domains. In particular, this work aims to contribute to the next generation of clinical systems that can 1) scale to include many millions of patient records and 2) use a single, distributed patient representation to effectively support clinicians in their daily activities, rather than multiple systems working with different patient representations derived for different tasks \cite{miotto2016deep}. 

To this aim, enabling efficient data-driven patient stratification analyses to identify disease subgroups is an important aspect to unlock personalized healthcare. Ideally, when new patients enter the medical system, their health status progression can be tied to a specific subgroup, thereby informing the treating clinician of personalized prognosis and possible effective treatment strategies, or counseling in cases where a certain diagnosis is difficult and a more thorough examination is required (e.g, specific genetic or lab tests). Moreover, the clinical characteristics of the different subtypes can potentially lead to intuitions for novel discoveries, such as comorbidities, side-effects or repositioned drugs, which can be further investigated analysing the patient clinical trajectories. 

Previous studies mostly focused on a specific disease using ad hoc cohorts of patients and features \cite{baytas2017patient, doshivelez2013, li2015identification, zhang2019data, lombardo2016, stevens2019}. While these studies obtained relevant clinically meaningful results, the computational framework is hard to replicate for different diseases and it is tied to the specific study and to the specific data. Deep learning has extensively been used to model EHRs for medical analysis \cite{miotto17review, xiao2018review}, including clinical prediction, such as disease onset, mortality, and readmission \cite{choi15, deepcare2016, rajkomar_scalable_2018}, and disease phenotyping \cite{miotto2016deep, beaulieu2016semi}. Because deep learning methods have not yet been leveraged for disease subtyping at scale, ConvAE aims to fill this gap and to provide an architecture that can improve unsupervised EHR pre-processing to favor patient stratification and unveil clinically meaningful and actionable insights. Additionally, unlike previous representation learning methods which did not consider the temporality of EHRs \cite{miotto2016deep, beaulieu2016semi}, ConvAE uses CNNs in combination with embeddings to specifically capture some of the longitudinal aspects of patient clinical status, leading to more robust representations. CNNs were already used to model EHRs for specific predictive analysis, as part of supervised architectures \cite{nguyen2017deepr, suo2018deep}. Differently, we trained CNNs in an unsupervised framework based on autoencoders to learn general-purpose patient representations. While these representations were used to leverage disease subtype discovery, they can also be fine-tuned and applied to specific supervised tasks, such as disease phenotyping and prediction.

There are several limitations to our study. First, we acknowledge that the lack of any discernible pattern in the multi-disease clustering analysis can also be due to noise and biases in the data, which might affect both learned representations and clustering. In particular, processing EHRs with minimum data engineering, on the one hand, preserves all the available information and, to some extent, prevents systematic biases. On the other, it adds hospital-specific biases intrinsic to the EHR structure and noise due to data being redundant and too generic. Improving EHR pre-processing by, e.g., better modeling clinical notes and/or improving feature filtering, should help reduce noise and improve performances. Second, we identified patients related to complex disorders using SNOMED--CT codes and this likely led to the inclusion of many false positives that affected the learning algorithms \cite{wei2015}. The use of phenotyping algorithms based on manual rules, e.g., PheKB \cite{kirby2016}, or semi-automated approaches, e.g. \cite{halpern2016, glicksberg2017}), should help identify better cohorts of patients and, consequently, better disease subtypes. Another limitation comes from the choice, among all possibilities, of the specific complex disorders. This allowed us to test the approach on heterogeneous conditions that affect different biological mechanisms, showing the efficacy of the proposed framework in generalizing to various clinical domains. Nevertheless, the approach should be further evaluated with other typologies of conditions as well, such as multiple sclerosis, autoimmune diseases, and psychiatric disorders. Lastly, we identified relevant concepts in the patient subgroups by simply evaluating their frequency. Adding a semantic modeling component based on, e.g., topic modeling \cite{blei2003} or word embeddings \cite{mikolov2013efficient}, might lead to more clinically meaningful patterns.

Future works will attempt to address these limitations and to further improve and replicate the architecture. First, we plan to enable multi-level clustering in order to stratify patients within the subtypes. This should lead to more granular patient stratification and thus, to patterns on a more individual-level. Second, we plan to verify ConvAE generalizability by replicating the study on EHRs from different healthcare institutions. Third, we will evaluate the use of disease subtypes as labels for training supervised models that can predict stratified patient risk scores. This, beside further validating the relevance of the results, will also provide an initial and intuitive framework to apply the results of patient stratification to clinical practice. To this aim, we plan to first assess treatment safety and efficacy between subtypes of a specific disease. Finally, to develop more comprehensive disease characterizations, we will include other modalities of data, e.g., genetics, into this framework, which will hopefully refine clustering and reveal new etiologies. Multi-modal stratified disease cohorts promise to facilitate better predictive capabilities for future outcomes by modeling how molecular mechanisms interact with clinical states.

\section*{Methods}

The framework to derive patient representations that enable stratification analysis at scale is based on 3 steps: 1) data pre-processing; 2) unsupervised representation learning (i.e., ConvAE); and 3) clustering analysis of disease-specific cohorts (see Figure~\ref*{fig:arch}a). In this section, we report details of this framework as well as the description of the evaluation design.

\subsection*{Dataset}

We used de-identified EHRs from the Mount Sinai Health System data warehouse; the study was approved by IRB-19-02369 in accordance with HIPAA guidelines. Mount Sinai Health System is a large and diverse urban hospital located in New York, NY, which generates a high volume of structured, semi-structured and unstructured data from inpatient, outpatient, and emergency room visits. Patients in the system can have up to $12$ years of follow-up data unless they are transferred or move their residence away from the hospital system. We accessed a de-identified dataset containing approximately $4.2$ million patients, spanning the years from $1980$ to $2016$.

For each patient, we aggregated general demographic details (i.e., age, sex, and race) and clinical descriptors. We included ICD-9 diagnosis codes, medications normalized to RxNorm, CPT-4 procedure codes, vital signs, and lab tests normalized to LOINC. ICD-10 codes were mapped-back to the corresponding ICD-9 versions. We pre-processed clinical notes using a tool based on the Open Biomedical Annotator that extracts clinical concepts from the free-text \cite{jonquet2009, lependu2012}. The vocabulary $V$ was composed by $57,464$ clinical concepts. 

We retained all patients with at least two concepts, resulting in a collection of $1,608,741$ different patients, with an average of $88.9$ records per patient. In particular, the cohort included $900,932$ females, $691,321$ males, and $16,488$ not declared; the mean age of the population as of $2016$ was $48.29$ years ($sd=23.79$). Patients were randomly partitioned in half for 2-fold cross-validation to assess model generalizability and replicability of the results. In each train set, we retained $30,000$ random patients for tuning the model hyperparameters. Train and test pre-processed sets' details are reported in Supplementary Table~\ref*{tab:stats}.

\subsection*{Data pre-processing}

Every patient in the dataset is represented as a longitudinal sequence $s_p$ of length $M$ of aggregated temporally-ordered medical concepts, i.e., $s_p = (w_1,\;w_2,\;\dots,w_{M})$, where each $w_i$ is a medical concept from the vocabulary $V$. Pre-processing includes: 1) filtering the least and most frequent concepts; 2) dropping redundant concepts within fixed time frames; 3) splitting long sequences of records to include the complete patient history while leveraging the CNN framework, which requires fixed-size inputs.

We consider all the EHRs as a document $D$ and each patient sequence $s_p$ as a sentence. For each concept $w$ in $V$ we first compute the probability of having $w$ in $D$. We then multiply this by the sum of the probabilities to find $w$ in a sentence $s_p$ for all sentences. In particular, let $P$ be the set of all patients, $\forall w \in V$, the filtering score is defined as:

\begin{equation}
\label{eq:filter}
\mathbb{P}(w \in D)\sum_{p\in P}\mathbb{P}(w \in s_p)=\frac{\# \{s\in D;\ w\in s\}}{|D|}\sum_{p\in P}\frac{\#\{w_i \in s_p;\ w_i = w \}}{|s_p|}, 
\end{equation}

where $|D|$ is the total number of sentences and $|s_p|$ is the length of a patient sequence. The filtering score combines document frequency, i.e., number of patients with at least one occurrence of $w$, and term frequency, i.e., total number of occurrences of $w$ in a patient sequence. We then drop all concepts with filtering scores outside certain cut-off values to reduce the amount of noise (i.e., not informative concepts that occur multiple times in few patients, or too general concepts that occur in many patients).

A patient may have multiple encounters in their health records that span consecutive days and might include repeated concepts that are often artifacts of the EHR system, rather than new clinical entries. To reduce this bias, we drop all duplicate medical concepts from the patient records within overlapping time intervals of $T$ days. Within the same time window, we also randomly shuffle the medical concepts, given that events within the same encounter are generally randomly recorded \cite{choi2016learning, glicksberg2017}. Lastly, we eliminate all patients with less than $3$ concepts in their records.

Patient sequences are then chopped into subsequences of fixed length $L$ that are used to train the ConvAE model. Each patient sequence is thus defined as:

\[
s_p=[(w_1, \dots, w_L), (w_{L+1}, \dots, w_{2L}), \dots],
\]

and subsequences shorter than $L$ are padded with $0$ up to length $L$. For the sake of clarity, in the following section we present the architecture as applied to a general subsequence $s = (w_1, \dots, w_L)$.

\subsection*{The ConvAE architecture}

ConvAE is a representation learning model that transforms patient EHR subsequences into low-dimensional, dense vectors. The architecture consists of three stacked modules (see Figure \ref*{fig:arch}b). This study proposes to use in combination embedding, CNNs, and autoencoders to process EHRs and to derive unsupervised vector-based patient representations that can be used for clinical inference and medical analysis.

Given $s$, the architecture first assigns each medical concept $w$ to an $N$-dimensional embedding vector $v_w$ to capture the semantic relationships between medical concepts. Specifically, a patient subsequence is represented as an $(L\times N)$ matrix $E = (v_{w_1},\;v_{w_2},\;\dots,v_{w_L})^T$, where $L$ is the subsequence length, and $N$ is the embedding dimension. This structure also retains temporal information because the rows of matrix $E$ are temporally ordered according to patient visits.

The architecture is then composed by CNNs, which extract local temporal patterns, and AEs, which learn the embedded representations for each patient subsequence. The CNN applies temporal filters to each embedding matrix. CNN filters applied to EHRs usually perform a one-side convolution operation across time via filter sliding. A filter can be defined as $k \in \mathbb{R}^{h\times N}$, where $h$ is the variable window size and $N$ is the embedding dimension \cite{zhu2016, suo2017personalized}. Our approach differs in that it processes embedding matrices as they were RGB images carrying a third ``depth'' dimension. With this approach, we enable the model filters to learn independent weights for each encoding dimension, thus activating for the most salient features in each dimension of the embedding space. Therefore, we reshape the $(L\times N)$ embedding matrix into $\tilde{E} \in \mathbb{R}^{1\times L \times N}$ and we consider the embedding dimensions as channels. We then apply $f$ filters $\mathbf{k} \in \mathbb{R}^{1\times h\times N}$ to the padded input to keep the same output dimension and learn features that may grasp sequence characteristics. In particular, for each filter $j$, we obtain:

\begin{equation}
(R)_j=\relu(\sum_{i=0}^{N-1}\mathbf{k}_i\star\tilde{\mathbf{e}}_i + \mathbf{b}_j), \ j=1,\dots,f,
\label{eq:cnn}
\end{equation}

where: $R \in \mathbb{R}^{1\times L\times f}$ is the output matrix; $\mathbf{k}_i$ is the $h$-dimensional weight matrix at depth $i$; $\tilde{\mathbf{e}}_i \in \mathbb{R}^{1\times L}$ is the $i$-th embedding dimension of the input matrix; $\mathbf{b}$ is the bias vector; and $(\star)$ is the convolution function. We used Rectified Linear Unit (ReLU) as the activation function and max pooling. The output is then reshaped into a concatenated vector of dimension $L\cdot f$. This configuration learns different weights for each embedding dimension to highlight relevant interdependencies of medical concepts, and tune representations of patient histories to identify the most relevant characteristics of their semantic space. 

We then use fully dense layers of autoencoders to derive embedded patient representations that estimate the given input subsequences. Specifically, we extract the hidden representation $\mathbf{y}$, a $H$-dimensional vector, as the encoded representation of each patient subsequence. Each patient sequence $s_p$ is then transformed into a sequence of encodings $s_h$ that can be post-modeled to obtain a unique vector-based patient representation. Here we simply component-wise average all the subsequence representations.

To train ConvAE, we set up a multi-class classification task that reconstructs each initial input one-hot subsequence of medical terms, from their encoded representations. Given a subsequence of medical concepts $s$, the ConvAE is trained by minimizing the Cross Entropy (CE) loss:

\begin{displaymath}
\crossentropy(\softmax(O),\;s) = -\frac{1}{L}\sum_{j=1}^{L}\log(\softmax(O^{j})_{w_{j}}),
\end{displaymath}

where $O$ is the output of ConvAE reshaped into a matrix of dimension $|V| \times L$, $w_{j}$ is the $j$-th element of sequence $s$ that correspond to a term indexed in $V$ and:

\begin{equation}
\softmax(O^j)_i=\frac{\exp{O^j_i}}{\sum_{i=1}^{|V|}\exp{O^j_i}}\ \ \  i=1,...,|V|.
\label{softmax}
\end{equation}

Since the objective function consists of only self-reconstruction errors, the model can be trained without any supervised training samples.

\subsection*{Clustering analysis for patient stratification}

ConvAE-based representations can be used to stratify patients from any preselected cohort without needing additional feature engineering or manual adjustments. To this aim, patients with a specific disease are selected using, e.g., ICD codes, SNOMED--CT diagnosis, or phenotyping algorithms (e.g., \cite{wei2015, halpern2016, glicksberg2017}), and clustering is applied to the corresponding representations to identify disease subgroups. Here, specifically, we use SNOMED--CT diagnosis to preselect the disease cohorts and hierarchical clustering with Ward's method and Euclidean distance to derive disease subgroups. We identify the number of subclusters that best disentangles heterogeneity on the disease dataset using the Elbow Method, which empirically selects the smallest number of clusters that minimize the increase in explained variance.

A systematic analysis of the patients in each subgroup can then automatically identify the medical concepts that significantly and uniquely define each disease subtype. In this work, we rank all the codes by their frequency in the patient sequences. In particular, we compute the percentages of patients whose sequence includes a specific concept both with respect to a subcluster (i.e., in-group frequency) and to the complete disease cohort (i.e., total frequency). Ranking maximizes, first, the in-group percentage, and second, the total percentage. We then analyze the most frequent concepts and we use a pairwise chi-squared test to determine whether the distributions of present/absent concepts with respect to the detected subgroups are significantly different \cite{zhang2019data}.

\subsection*{Implementation details}

All model hyperparameters were empirically tuned to minimize the network reconstruction error, while balancing training efficiency and computation time. We tested a large amount of configurations (e.g.,  time interval $T$ equal to $\{15,\ 30\}$; patient subsequence length $L$ equal to $\{32,\ 64\}$; embedding dimension $N$ spanning $\{100,\ 128,\ 200\}$). For brevity, we report only the final setting used in the patient stratification experiments. All modules were implemented in Python $3.5.2$, using \texttt{scikit-learn} and \texttt{pytorch} as machine learning libraries \cite{scikit-learn, paszke2017}. Computations were run on a server with an Nvidia Titan V GPU.

We used equation \eqref{eq:filter} to discard terms with a filtering score less than $10^{-6}$, i.e., document frequency ranging from $1$ to $10$. Examples of discarded concepts are \emph{clotrimazole}, an antifungal medication, and \emph{torsemide}, a medication to reduce extra fluid in the body. We decided to retain all the very frequent concepts as most of them seemed clinically informative (e.g., vital signs). Patients with less than $3$ medical concepts were then discarded. In total, $24,665$ medical terms were filtered out, decreasing the vocabulary size to $32,799$.

We divided each patient history in consecutive, half-overlapped temporal windows of $T=15$ days, shuffled unique medical concepts and dropped redundant terms. Patient sequences were then split in subsequences of length $L=32$ concepts, obtaining about $\sim3M$ subsequences of medical concepts for training. This value was chosen to enable efficient training of the autoencoder with GPUs.

We initialized medical concept embeddings using word2vec with the skip-gram model \cite{mikolov2013efficient}. We considered all the subsequences in the training set as sentences and medical concepts as words \cite{glicksberg2017, choi2016learning}. We obtained $100$-dimensional embeddings for $31,659$ medical concepts of the vocabulary. The remaining concepts were initialized randomly; the subsequence padding was initialized as the null vector (i.e., at $\mathbf{0}$). These embedding vectors were then used as input for the ConvAE module and were further refined during the model training.

The CNN module used 50 filters with kernel size equal to 5 and $\relu$ activation function. The autoencoder was composed by 4 hidden layers with 200, 100, 200 and $|V| \times 32$ hidden nodes, respectively, where $|V|$ is the vocabulary size. We used $\relu$ activation in the first three layers and $\softplus$ activation in the final layer to obtain continuous output. We applied dropout with $p=0.5$ in the first two layers for regularization. The model was trained using cross entropy loss with the Adam optimizer (learning rate = $10^{-5}$ and weight decay = $10^{-5}$) \cite{kingma2014} for 5 epochs on all training data and batch size of $128$. The size of the patient representations was equal to $100$.
 
We evaluated different CNN configurations composed by 1-layer (i.e., ``ConvAE 1-layer CNN''), 2-layers (i.e., ``ConvAE 2-layer CNN''), and one multikernel layer (i.e., ``ConvAE multikernel CNN''). All hyperparameters were the same, except the number of filters in the second CNN of the 2-layer configuration that was set to $25$. Multikernel CNN performs parallel training of distinct CNNs with different kernel sizes, and concatenates the final outputs. We used kernel dimensions equal to $3$, $5$, and $7$.

\subsection*{Baselines}

We compared ConvAE with the following representation learning algorithms: ``RawCount'', ``SVD-RawCount'', ``SVD-TFIDF'', and ``Deep Patient''. All baselines derived vector-based patient encodings of size $100$.

RawCount is a sparse representation where each patient is encoded into a count vector that has the length of the vocabulary. More specifically, each individual health history $s_p$ is represented as an integer vector $\mathbf{x} \in \mathbb{Z}^{|V|}$, where each element is the frequency of the corresponding clinical concept in the patient longitudinal history , i.e., $x_i=\#\{w_i;\ w_i \in s_p\}$. 

SVD-RawCount applies truncated singular value decomposition (SVD) to the RawCount matrix to compute the largest singular values of the raw count encodings, which define the dense, lower-dimensional representations.

SVD-TFIDF transforms the raw count encodings using the term frequency–inverse document frequency (TFIDF) weighting schema and applies truncated SVD to the resulting matrix. We considered the patient EHR sequences as documents, the entire dataset as corpus and we derived TFIDF scores for all medical concepts. Each patient is then represented as a vector of length $|V|$, with the corresponding TFIDF weight for each concept, and the matrix obtained is reduced via truncated SVD.
    
Deep Patient transforms the raw count matrix using a stack of denoising autoencoders as proposed by Miotto et al. \cite{miotto2016deep}. We used the implementation details presented in the paper, with batch size equal to $32$, corruption noise equal to $5\%$, and 5 training epochs.
   
\subsection*{Multi-disease clustering analysis}

We evaluated all the representation learning approaches in a clustering task to determine how they were able to disentangle patients with different conditions. We chose eight complex disorders: type 2 diabetes (T2D), multiple myeloma (MM), Parkinson's disease (PD), Alzheimer's disease (AD), Crohn's disease (CD), prostate cancer (PC), breast cancer (BC) and attention deficit hyperactivity disorder (ADHD). We retrieved all the corresponding patients in the test sets using SNOMED--CT codes after verifying that at least one correspondent ICD-9 code was present in a patient EHRs. In particular, we looked for \emph{Type 2 diabetes mellitus (250.00)} for T2D; \emph{Multiple myeloma without mention of having achieved remission (203.00)} for MM; \emph{Paralysis agitans (332.0)} for PD; \emph{Alzheimer's disease (331.0)} for AD; \emph{Regional enteritis of unspecified site (555.9)} for CD; \emph{Malignant neoplasm of prostate (185)} for PC; \emph{Malignant neoplasm of female breast (174.9)} for BC; and \emph{Attention deficit disorder with hyperactivity (314.01)} for ADHD. We discarded all patients with comorbidities within the selected diseases to facilitate the clustering interpretation. We then performed hierarchical clustering with $k = 8$ clusters (i.e., same as the different diseases) for all the representations to evaluate if patients with the same condition were grouping together. The final test sets were composed by about $94,000$ patients per fold but were unbalanced, with disease cohorts ranging from about $1,900$ to $50,000$ patients (see Supplementary Table \ref*{tab:disclass}). To use balanced datasets and improve the efficacy of the experiment, we sub-sampled $5,000$ random patients for the highly populated diseases, and we iterated this subsampling process $100$ times, obtaining $100$ different clustering per test set.

We used entropy and purity scores averaged across the $100$ experiments of each fold to measure to what extent the clusters matched the different diseases. In particular, for each cluster $j$, we define the probability that a patient in $j$ has disease $i$ as:

\begin{equation}
p_{ij}=\frac{m_{ij}}{m_j},
\end{equation}

where $m_j$ is the number of patients in cluster $j$ and $m_{ij}$ is the number of patients in cluster $j$ with a diagnosis of disease $i$. Entropy for each cluster is defined as: 

\begin{equation}
E_j=-\sum_{i}p_{ij}\log_2p_{ij},
\end{equation}

and conditional entropy $H(\text{disease}|\text{cluster})$ is then computed as:

\begin{displaymath}
    H(\text{disease}|\text{cluster}) = \sum_j\frac{m_j}{m}E_j,
\end{displaymath}

where $m$ is the total number of elements in the complex disease dataset.

Purity identifies the most represented disease in each cluster. For a cluster $j$, purity $P_j$ is defined as $P_j=\max_i p_{ij}$, where $p_{ij}$ is computed as before. The overall purity score is then the weighted average of $P_j$ for each cluster $j$. The perfect clustering obtains averaged entropy and purity scores equal to $0$ and $1$, respectively.

\subsection*{Disease subtyping analysis}

We evaluated the usability of ConvAE representations to discover disease subtypes for different and diverse conditions (i.e., patient stratification at scale). In particular, we selected a cohort of patients with T2D, PD, AD, MM, PC, and BC and ran hierarchical clustering on the ConvAE-based patient representations. These are all age-related complex disorders with late onset (i.e., increased prevalence after $60$ years of age \cite{cowie2018, lau2006, qiu2009, kazandjian2016, pc1, bc2}). We focused only on these conditions to attempt reducing confounding age effects that could affect the analysis of the subtypes (as it could happen on CD and ADHD cohorts, where a common onset age is less defined). To reduce noise in the sequence encodings, we averaged all patient subsequence representations from the first diagnosis forward, and we dropped sequences shorter than $3$ concepts. We ranged the number of clusters from $2$ to $15$ and we used the Elbow Method to empirically select the smallest number of clusters that minimize the increase in explained variance. We then performed a qualitative analysis of each subtype, similarly to Zhang et al. \cite{zhang2019data}, to identify which medical concepts characterized the specific group of patients. We further verified the various subgroups in the medical literature and with the support of a practicing clinician.

\section*{Data availability}
The data used for this study are available from the Mount Sinai Health System (NYC), but restrictions apply to the availability of these data, which were used under license for the current study, and so are not publicly available. Data are however available from the authors upon reasonable request and with permission of Mount Sinai Health System.

\section*{Code availability}
Code is available at: \url{https://github.com/landiisotta/convae_architecture}.

\section*{Acknowledgments}

R.M. would like to thank the support from the Hasso Plattner Foundation, the Alzheimer’s Drug Discovery Foundation and a courtesy GPU donation from Nvidia. I.L. acknowledges the support from the Bruno Kessler Institute.

\section*{Competing interests}

The authors declare no competing interests.

\section*{Author contributions}

I.L. and R.M. conceived and designed the work. I.L. conducted the research and the experimental evaluation, and drafted the manuscript. R.M. created the dataset, supervised and supported the research, and substantially edited the manuscript. B.S.G. substantially edited the manuscript and created the architecture figures. H.L. and S.C. advised on methodological choices and critically revised the manuscript. G.L. provided clinical validation of the results and critically revised the manuscript. M.D. revised the manuscript and contributed to the interpretation of the data. J.T.D. and C.F. supported the research and revised the manuscript. All the authors gave final approval of the completed manuscript version and are accountable for all aspects of the work.

\newpage
\singlespacing
\setlength\bibitemsep{8pt}
\printbibliography[title={References}]

\clearpage
 
  \begin{table}
 \ra{1.3}
 \begin{tabular}{@{}rccc@{}}\toprule
 & \textbf{Entropy$^1$} & \textbf{Purity$^1$} & \textbf{Disease Number$^2$}\\ \midrule
 ConvAE 1-layer CNN & $2.61$ ($0.04,\ [2.58, 2.67]$)$^{***}$ & $0.31$ ($0.02,\ [0.31, 0.35]$)$^{***}$ & $6.50$ ($0.62$)$^{***}$\\
 ConvAE 2-layer CNN & $2.75$ ($0.02,\ [2.74, 2.78]$) & $0.26$ ($0.01,\ [0.26, 0.29]$) & $5.93$ ($0.50$)\\
 ConvAE multikernel CNN & $2.66$ ($0.03,\ [2.64, 2.70]$) & $0.30$ ($0.02,\ [0.29, 0.33]$) & $5.94$ ($0.47$)\\ \midrule
 RawCount & $2.90$ ($0.02,\ [2.88, 2.92]$) & $0.18$ ($0.01,\ [0.18, 0.20]$) & $4.76$ ($0.70$)\\
 SVD-RawCount & $2.90$ ($0.01,\ [2.90, 2.92]$) & $0.19$ ($0.01,\ [0.18, 0.20]$) & $5.13$ ($0.79$)\\
 SVD-TFIDF & $2.85$ ($0.02,\ [2.84, 2.87]$) & $0.21$ ($0.01,\ [0.21, 0.23]$) & $5.83$ ($0.76$)\\
 Deep Patient & $2.81$ ($0.02,\ [2.80, 2.84]$) & $0.24$ ($0.01,\ [0.23, 0.25]$) & $5.96$ ($0.74$)\\
 \bottomrule
 \multicolumn{4}{l}{\footnotesize $^1$ Mean (sd, CI); $^2$ Mean (standard deviation); $^*p<0.05;\ 
  ^{**} p<0.01;\ ^{***} p<0.001$; }\\
 \multicolumn{4}{l}{\footnotesize CNN = Convolutional Neural Network; SVD = Singular Value Decomposition;}\\
 \multicolumn{4}{l}{\footnotesize TFIDF = Term Frequency - Inverse Document Frequency}
 \end{tabular}
 \caption{Multi-disease clustering performances of ConvAE configurations and baselines. The scores reported are averaged over a 2-fold cross-validation experiment. ConvAE 1-layer CNN significantly outperforms all other configurations and baselines on all measures. Multiple pairwise t-tests with Bonferroni correction are used to compare performances.}
 \label{tab:result}
 \end{table}
  
  \clearpage
  
 \clearpage
 
 \includepdf[pages=-]{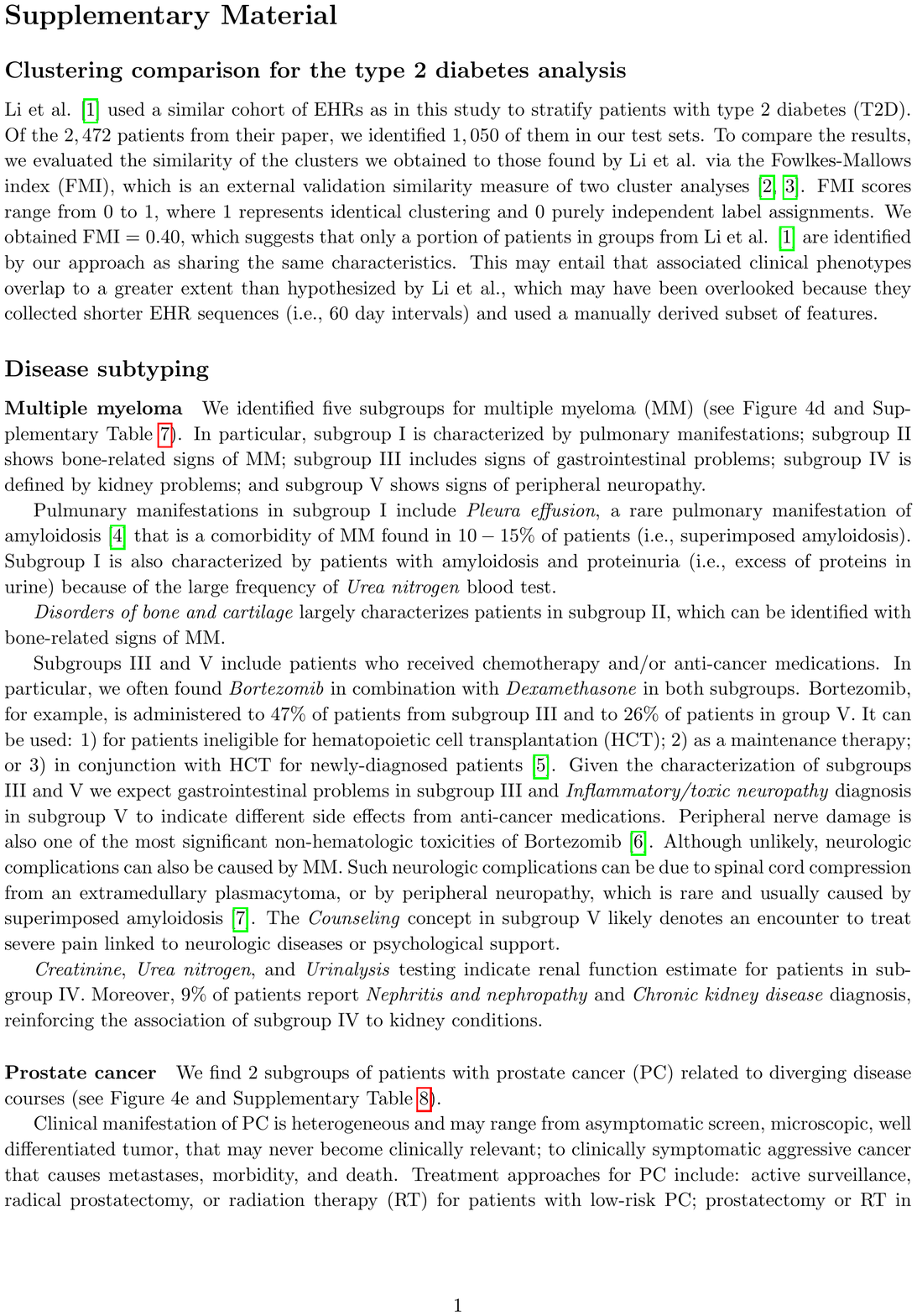}
 
\end{document}